\documentclass[final,3p]{elsarticle}

\usepackage{graphicx}

\usepackage{alltt}

\usepackage{amsmath,amssymb}

\usepackage{yhmath} 
\usepackage{srctex} 
\usepackage[dvipsnames]{color}

\usepackage{url}

\usepackage{yhmath}

\usepackage{srctex} 

\def\m#1{\mathsf{#1}} 

\newcommand{\cl}[2]{\ensuremath{\mathit{Cl}_{#1,#2}}}

\DeclareMathOperator{\Det}{Det} 

\DeclareMathOperator{\abs}{abs}

\DeclareMathOperator{\arctanh}{arctanh}

\newcommand{\bbR}{\ensuremath{\mathbb{R}}}

\newcommand{\bbZ}{\ensuremath{\mathbb{Z}}}

\newcommand{\reverse}[1]{\widetilde{#1}}
\newcommand{\gradeinverse}[1]{\wideparen{#1}}
\newcommand{\cliffordconjugate}[1]{\widetilde{\wideparen{#1}}}

\newcommand{\ii}{\mathrm{i}}
\newcommand{\jj}{\mathrm{j}}
\newcommand{\kk}{\mathrm{k}}
\newcommand{\ee}{\mathrm{e}} 

\def\m#1{\mathsf{#1}}

\def\e#1{\mathbf{e}_{#1}} 

\newcommand{\ba}{\ensuremath{\mathbf{a}}}
\newcommand{\bb}{\ensuremath{\mathbf{b}}}

\newcommand{\bv}{\ensuremath{\mathbf{v}}}

\newcommand{\cA}{\ensuremath{\mathcal{A}}}
\newcommand{\cB}{\ensuremath{\mathcal{B}}}

\renewcommand{\d}{\ensuremath{\mathbin{\cdot}}} 
\newcommand{\w}{\ensuremath{\mathbin{\wedge}}} 

\usepackage{color}
\newcommand{\blue}[1]{\textcolor{blue}{#1}}
\newcommand{\red}[1]{\textcolor{red}{#1}} 
\newcommand{\mycomment}[1]{} 

 \numberwithin{equation}{section}

\newcommand{\magnitude}[1]{\lvert #1\rvert}

\usepackage{color}

\begin{document}

\begin{frontmatter}

\title{Exponential and logarithm of multivector in low
dimensional ($n=p+q<3$) Clifford algebras}

\author[mymainaddress,mysecondaryaddress]{Adolfas Dargys}
\ead{adolfas.dargys@ftmc.lt}

\address[mymainaddress]{Center for Physical Sciences and Technology, Semiconductor
Physics Institute, Saul{\.e}tekio 3, LT-10257
Vilnius, Lithuania}

\author[mymainaddress2,mysecondaryaddress2]{Art{\=u}ras Acus}
\ead{arturas.acus@tfai.vu.lt}

\address[mymainaddress2]{Institute of Theoretical Physics and Astronomy, Vilnius
University, Saul{\.e}tekio 3, LT-10257 Vilnius, Lithuania}

\begin{abstract}
Closed form expressions for a multivector exponential and
logarithm  are presented in real Clifford geometric algebras
$\cl{p}{q}$ when $n=p+q=1$ (complex and hyperbolic numbers) and
$n=2$ (Hamilton, split and conectorine quaternions). Starting from
\cl{0}{1} and \cl{1}{0} algebras wherein square of a basis vector
is either $-1$ or $+1$, we have generalized exponential and
logarithm formulas to 2D quaternionic algebras, \cl{0}{2},
\cl{1}{1}, and \cl{2}{0}. The sectors in the multivector
coefficient space where 2D logarithm exists are found. They are
related with a square root of the multivector.
\end{abstract}

\end{frontmatter}










\section{Introduction\label{sec:intro}}
Quaternion algebras find  a wide application in graphics, robotics
and control of spatial rotation of solid  bodies, including
aerospace flight
dynamics~\cite{Kuipers99,Adler95,Gurlebeck1997,Morais14}. During
the last ten years there is a tendency to replace quaternions by
multivectors (MVs)  of geometric (aka Clifford) algebras (GAs),
mainly due to the possibility to carry out calculations in higher
dimensional GAs of  mixed
signatures~\cite{Hanson2006,Macdonald2010,Macdonald2012,Girard07,Lavor2018}
and, consequently  to employ wider GA  capabilities. Of special
mention is conformal \cl{4}{1} GA that allows to do complicated
graphics in 5D vector space and then transform the graphics to 3D
Euclidean space for visualization~\cite{Perwass09}.

In this paper we investigate low dimensional algebras from GA
perspective, namely, 1D complex and hyperbolic number algebras as
well as 2D algebras \cl{0}{2}, \cl{1}{1} and \cl{2}{0} that are
isomorphic to quaternionic algebras: the Hamilton quaternion (or
briefly the quaternion) \cite{Gurlebeck1997,Morais14},
coquaternion also known as a split
quaternion~\cite{Ozdemir2009,Falcao2018}, and
conectorine~\cite{Opfer2017}. The properties of the Hamilton
quaternion,  which is isomorphic to \cl{0}{2}, recently have been
summarized in a handbook~\cite{Morais14}. The coquaternion and
conectorine are less known. They are isomorphic to \cl{1}{1} and
\cl{2}{0} algebras that are noncommutative too. As we shall see,
in all 2D algebras the exponential and logarithm may be treated in
a uniform way if they are reformulated in GA terms what, in turn,
helps to generalize and better understand known properties as well
as to discover new ones, for example, continuous degrees of
freedom related to a free vector pointing in an arbitrary
direction~\cite{AcusDargys2022a}. In this paper, general formulas
for exponential and logarithm  are presented in a form of sum of
GA basis elements and in a basis-free form. The subject considered
in this paper is akin to exponential factorization of MV  into
product of exponentials~\cite{Hitzer2020a} and square root of
MV~\cite{Dargys2019,AcusDargysPreprint2020}.

In Sec.~\ref{sec:notations},  notation and general  properties of
GA exponential and logarithm functions are introduced. The GA
expressions  in 1D  are presented in Sec.~\ref{sec:ExpLog1D}. In
Secs.~\ref{sec:ExpLog2D} and \ref{sec:Log3D}, respectively, the
exponential and logarithm in 2D are considered. In Addendum
(Sec.~\ref{sec:sqrt}) the square root of MV is discussed. Finally,
in Sec.~\ref{sec:conclusion} the conclusion and short discussion
are given.

\section{Properties of exponential and logarithm in
GA\label{sec:notations}} Let $\e{i}$ be the basis vector  and
$\e{ij}\equiv\e{i}\e{j}=-\e{ji}$ be the bivector. The latter is
the geometric product of two orthogonal basis vectors. Complex and
hyperbolic numbers (aka Clifford numbers) in GA~\cite{Sobczyk2013}
are represented by the following MVs
\begin{equation}\label{complexHyperbolic}\begin{split}
\cl{0}{1},\quad&\m{A}=a_0+a_1 I,\quad\text{where\ }I\equiv\e{1}\ \text{and\ }\e{1}^2=-1,\\
\cl{1}{0},\quad&\m{A}=a_0+a_1 I,\quad\text{where\ }I\equiv\e{1}\ \text{and\ }\e{1}^2=+1,\\
\end{split}\end{equation}
where $a_0$ and $a_1$ are the real coefficients, $a_0,a_1\in\bbR$.
$a_0$ is called the scalar part of MV and $a_1 I$  the
pseudoscalar. In 1D GAs the basis vectors coincide with an
elementary pseudoscalar $I$. The squares,  $\e{1}^2\equiv
\ii^2=-1$ in \cl{0}{1}  and $\e{1}^2=1$ in \cl{1}{0}, suggest that
we have to do with complex and hyperbolic numbers, respectively.

In 2-dimensional (2D)  algebras there are two basis vectors
$\e{1}$ and $\e{2}$, and a bivector $\e{12}\equiv I$ (oriented
plane). The general Clifford number $\m{A}$ is
\begin{equation}\label{quaternonic}\begin{split}
&\cl{0}{2}:\ \m{A}=a_0+a_1\e{1}+a_2\e{2}+a_{12} I,\ \text{where\ }\ \e{1}^2=\e{2}^2=-1,\ I^2=-1,\\
&\cl{1}{1}:\ \m{A}=a_0+a_1\e{1}+a_2\e{2}+a_{12} I,\ \text{where\ }\ \e{1}^2=-\e{2}^2=-1,\ I^2=1,\\
&\cl{2}{0}:\ \m{A}=a_0+a_1\e{1}+a_2\e{2}+a_{12} I,\ \text{where\
}\ \e{1}^2=\e{2}^2=+1,\  I^2=-1.
\end{split}\end{equation}
The sum $\ba=a_1\e{1}+a_2\e{2}$ represents a general vector in 2D
bivector plane. The basis vectors satisfy $\e{1}\d\e{2}=0$
(orthogonality) and $\e{1}\w\e{2}=\e{1}\e{2}\equiv\e{12}$
(oriented unit plane), where the  dot and wedge denote the inner
and outer products. $\e{12}$  plays the role of an elementary
pseudoscalar~$I$. The sign of  $I^2$ depends on algebra,
Eq.~\eqref{quaternonic}. The algebras \cl{1}{1} and \cl{2}{0} are
isomorphic under the following exchange of GA basis elements:
$\e{1}\leftrightarrow\e{1}$, $\e{2}\leftrightarrow\e{12}$ and
$\e{12}\leftrightarrow\e{2}$.

The main  involutions, namely the reversion, inversion and
Clifford conjugation denoted, respectively, by tilde, circumflex
and their combination  are defined by the following changes in
component signs of MV $\m{A}=a_0+\ba+a_{12}I$,
\begin{equation}\begin{split}
&\widetilde{\m{A}}=a_0+\ba-a_{12}I,\qquad
\gradeinverse{\m{A}}=a_0-\ba+a_{12}I,\qquad
\cliffordconjugate{\m{A}}=a_0-\ba-a_{12}I.
\end{split}
\end{equation}
For complex and hyperbolic numbers there is only a single
involution, $\gradeinverse{\m{A}}=a_0+a_1\gradeinverse{I}=a_0-a_1
I$, that  usually is denoted  by asterisk in physics and
engineering and overline in mathematics.

The exponential of MV is another MV that  belongs to the same
geometric algebra \cl{p}{q}. If $\m{A}$ and $\m{B}$ are MVs, the
following properties hold:
\begin{equation}\label{expProperties}
\begin{split}
&\text{exp}(\m{A}+\m{B})=\text{exp}(\m{A})\text{exp}(\m{B})\quad\text{iff \ } \m{A}\m{B}=\m{B}\m{A},\\
&\widetilde{\ee^{\m{A}}}=\ee^{\widetilde{\m{A}}},\quad\gradeinverse{\ee^{\m{A}}}=\ee^{\gradeinverse{\m{A}}},
\quad \cliffordconjugate{\ee^{\m{A}}}=\ee^{\cliffordconjugate{\m{A}}},\\
\end{split}
\end{equation}
where $\ee$ is  the  base of the natural logarithm. In 1D algebras
the first property is always satisfied since the commutation of
scalar and vector is satisfied.

The GA exponential can be represented as a power series in a form
similar to scalar exponential~\cite{Lounesto97}. In numerical
form, i.e., when coefficients at basis elements $\e{1}$, $\e{1}$
and $\e{12}$ are real numbers, the exponential can be summed up
approximately~\cite{AcusDargys2022a}. To minimize the number of
multiplications it is convenient  to rewrite the exponential  in a
nested form (aka Horner's rule),
\begin{equation}\label{mercator}
\ee^{\m{A}}=1+\frac{\m{A}}{1}(1+\frac{\m{A}}{2}(1+\frac{\m{A}}{3}(1+\frac{\m{A}}{4}(1+\dots)))),
\end{equation}
which requires fewer MV products.  If numerical coefficients
in~$\m{A}$ are small enough the exponential $\ee^{\m{A}}$ can be
approximated by truncated series~\eqref{mercator} to high
precision. For examples, refer to paper~\cite{AcusDargys2022a}.

The following properties hold for MV logarithm:
\begin{equation}\begin{split}
&\text{log}(\m{A}\m{B})=\text{log}(\m{A})+\text{log}(\m{B})\quad\text{iff\ } \m{A}\m{B}=\m{B}\m{A},\\
&\ee^{-\text{log}(\m{A})}=\m{A}^{-1}, \\
&\widetilde{\log(\m{A})}=\log(\widetilde{\m{A}}),\quad\gradeinverse{\log(\m{A})}=\log(\gradeinverse{\m{A}}),\quad
\cliffordconjugate{\log(\m{A})}=\ee^{\cliffordconjugate{\m{A}}}.\\
\end{split}\end{equation}
When the logarithm of MV  exists, it may be approximated by series
\begin{equation}\begin{split}\label{logseries}
\log\m{B}=\m{B}(1+\m{B}(-\tfrac12+\m{B}(\tfrac13+\m{B}
(-\tfrac14+\m{B}(\tfrac15+\dotsm ))))),\quad0< \magnitude{\m{B}}<
1.
\end{split}\end{equation}
Here $|\m{B}|$ is the determinant
norm~\cite{AcusDargys2022a,AcusDargys2022b}. If logarithm exists,
a finite  series~\eqref{logseries} can be summed up  in a
numerical form approximately. However, as we shall see later there
may be sector(s) in the MV coefficient domain where the logarithm
does not exist at all.

In \cl{0}{1} algebra the norm $\magnitude{\m{B}}$, which is equal
to the square root of MV determinant
$\text{Det}(\m{B})=\m{B}\gradeinverse{\m{B}}=b_0^2+b_1^2>0$, is
called the magnitude or absolute value of the MV (or magnitude of
the complex number in this case). In hyperbolic number theory
similar role is played by product
$\m{B}\gradeinverse{\m{B}}=b_0^2-b_1^2$, which may be positive,
negative or zero.  In this case the magnitude called a determinant
semi-norm (or pseudonorm)
$\lVert\m{B}\rVert=\sqrt{\abs(\m{B}\gradeinverse{\m{B}})}=\sqrt{\abs(
b_0^2-b_1^2)}\ge 0$ is introduced. Note that now  the equality
sign appears, therefore,  the semi-norm $\lVert\m{B}\rVert$ may be
zero even if $\m{B}\ne 0$. The equality sign in case of the norm
$\magnitude{\m{B}}$ would require the MV to nullify.

\begin{figure}[t]
\centering
\includegraphics[width=5cm]{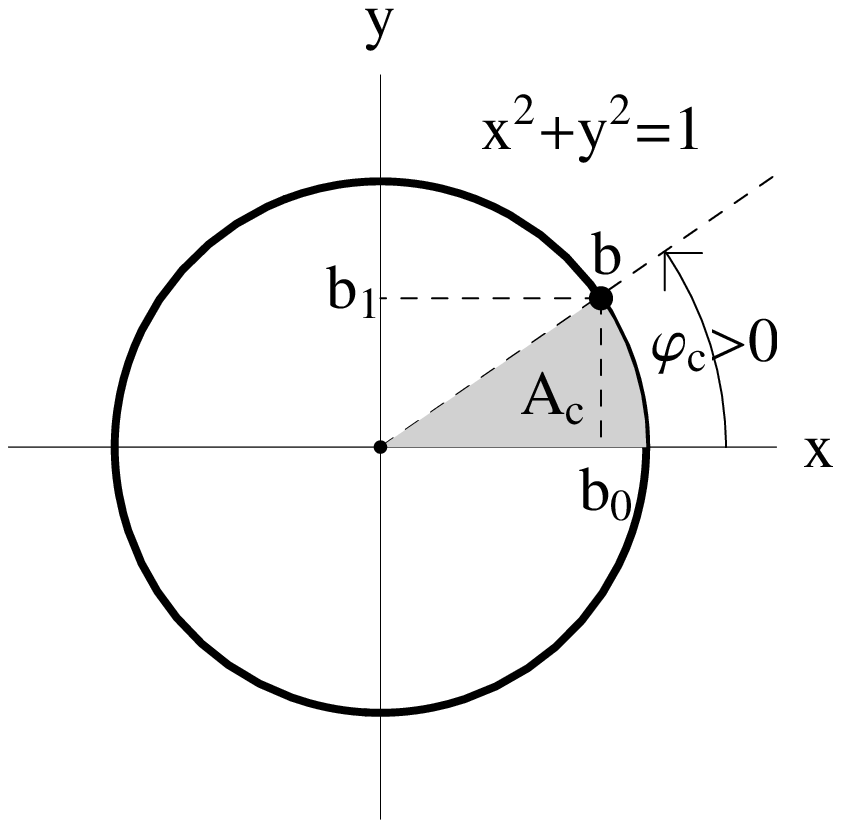}a)
\includegraphics[width=5cm]{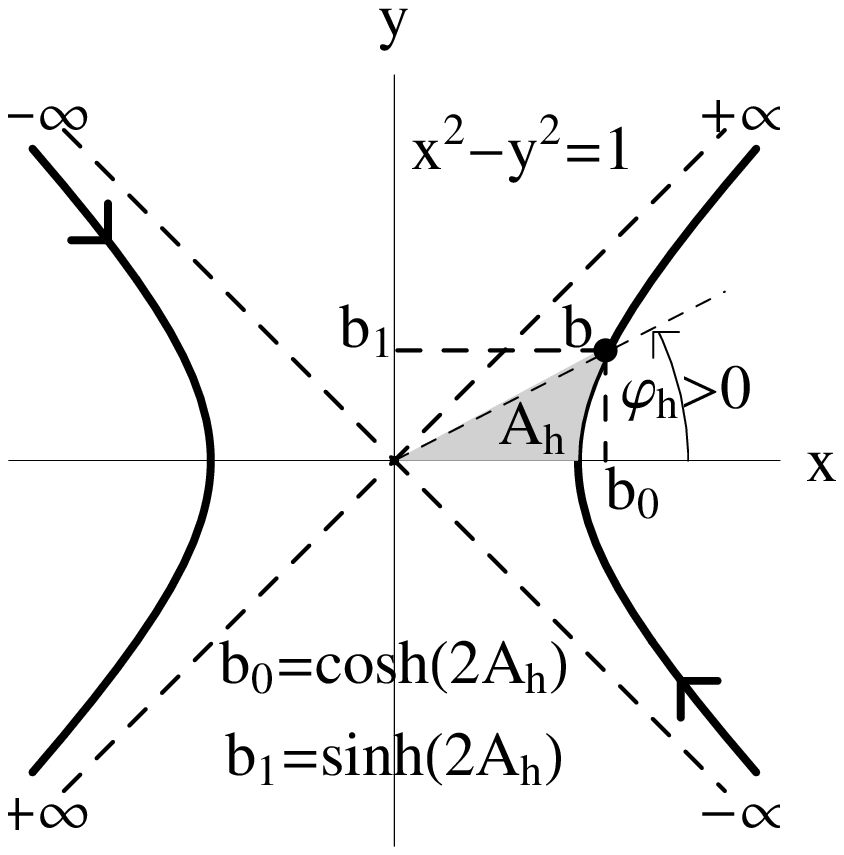}b)
\caption{ \label{fig:circeHyperbola}Analogy between unit circle
$x^2+y^2=1$ and unit hyperbola $x^2-y^2=1$.  For a circle the
coordinates of  point~$b$ are $x=\cos\varphi_c$ and
$y=\sin\varphi_c$, and for a hyperbola they are $x=\cosh\varphi_h$
and $y=\sinh\varphi_h$. The trigonometric and hyperbolic angles
are defined, respectively, by $\varphi_c=\arctan(y/x)$ in the
range $\varphi_c=[0,2\pi)$ and $\varphi_h=\textrm{artanh}(y/x)$ in
the range $\varphi_h=(-\infty,+\infty)$. For hyperbola they are
limited by asymptotes (dashed lines). The shaded areas $A_{c}$ and
$A_{h}$ are proportional to trigonometric and hyperbolic  angles:
$A_c=\varphi_c/2$ and $A_h=\varphi_h/2$. The infinity signs at
asymptotes  show extreme values of $\varphi_h/2$, where the
infinities having opposite signs meet,
$-\infty=+\infty$~\cite{Fenchel989}.}
\end{figure}

\section{Exponential and logarithm in 1D algebras\label{sec:ExpLog1D}}
One-dimensional GAs are represented by  two commutative algebras:
the well-known complex number algebra which is isomorphic to
\cl{0}{1} and the hyperbolic number algebra
\cl{1}{0}~\cite{Sobczyk2013}. In Fig.~\ref{fig:circeHyperbola} the
geometrical properties  of both algebras are compared graphically
on $x$-$y$ plane (equivalently on $b_0$-$b_1$  plane). In
Fig.~\ref{fig:circeHyperbola}b  the two branches of hyperbola
close down at plus/minus infinities~\cite{Fenchel989}. The shaded
area in both cases is proportional either to inner $\varphi_c$ or
outer $\varphi_h$ angle between the center and the point $b$ on
circle $y^2+x^2=1$ or hyperbola $y^2-x^2=1$, respectively. If a
point $b$ on the circle or hyperbola represents the MV
$\m{B}=b_0+b_1\e{1}$ then in GA the quantity
$\m{B}\gradeinverse{\m{B}}=b_0^2+b_1^2=\magnitude{\m{B}}^2>0$ is
the square of norm that graphically represents the sector $A_c$ in
Fig.~\ref{fig:circeHyperbola}a. Similarly the sector $A_h$ in
Fig.~\ref{fig:circeHyperbola}b represents the semi-norm
$\lVert\m{B}\rVert$ (pseudonorm) that as mentioned may be
positive, negative, or zero.

\subsection{Exponential and logarithm of MV in \cl{0}{1}}\label{logCL01}
Since \cl{0}{1} algebra is commutative and $\e{1}^2=-1$, we can
write
\begin{equation}\label{expCL01}
\ee^{\m{B}}=\ee^{b_0+b_1\e{1}}=\ee^{b_0}\ee^{b_1\e{1}}=\ee^{b_0}
\big(\cos b_1+\e{1}\sin b_1\big),
\end{equation}
where Euler's rule  was used. Presence of trigonometric functions
indicates that the exponential in \cl{0}{1} is a periodic function
with period $2\pi k$, where $k\in\bbZ$ is an arbitrary integer.
Thus, more generally  in \cl{0}{1} we have
$\ee^{\m{B}}=\ee^{b_0+b_1\e{1}+2\pi k\e{1}}$.

\begin{figure}[t]
\centering
\includegraphics[width=10cm]{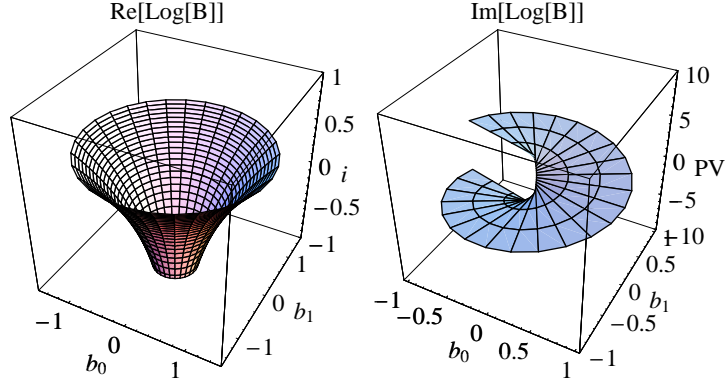}
\caption{ Graphical representation of \cl{0}{1} logarithm. The
real part
 $\text{Re}(\log(\m{B}))=\log(\sqrt{b_0^2+b_1^2})$ (left panel) and
principal logarithm $\varphi=\text{Im}(\log(\m{B}))$ in the range
$[-\pi,\pi]$ is represented by a single winding on the right
panel. At a fixed $\varphi$ the lines run parallel to horizontal
$\{b_0,b_1\}$ plane. \label{fig:PrincipalValue}}
\end{figure}

The logarithm of a complex number  $z=x+\ii y$  is
\begin{equation} \log z=\log(r\ee^{\ii
\varphi})=\log\magnitude{z}+\ii\varphi=\log\big(\sqrt{x^2+y^2}\big)+\ii\varphi,
\end{equation}
which in \cl{0}{1} algebra notation is
\begin{equation}\label{LogCL01}
\log\m{B}=\log\sqrt{b_0^2+b_1^2}+\e{1}\arctan(b_1/b_0)=\log\magnitude{\m{B}}+\e{1}\varphi.
\end{equation}
The angle $\varphi=\arctan(y/x)$, or $\varphi=\arctan(b_1/b_0)$,
is called the argument of logarithm. If $r=(b_0^2+b_1^2)^{1/2}$ is
const. then $\varphi$ may be interpreted as  a rotation angle of a
vector around coordinate center, Fig.~\ref{fig:PrincipalValue}a.
To eliminate sign ambiguity between quadrants \textit{1} and
\textit{3} (or \textit{2} and \textit{4}), the arc tangent of a
single argument usually is replaced by double argument arc tangent
$\arctan(x,y)$. If signs of  $x$ and $y$ are already fixed then
$\arctan(x,y)=\arctan(y/x)$. To include multiple rotations, after
every single rotation the period $2\pi$ is added  to $\varphi$, so
that after $k$ rotations we have $k$-windings in
Fig.~\ref{fig:PrincipalValue}b and angle
$\varphi=\text{arctan}(b_0,b_1)+2\pi k$, where
$k\in\bbZ=\dots-2,-1,0,1,2,\dots$. Similarly, in case of
hyperbolic functions to include the sign of $x$ and $y$ in the
quadrants \textit{1-4}  one may introduce a double angle
hyperbolic tangent\footnote{Figure~\ref{fig:circeHyperbola}b
represents properties of hyperbola drawn  on the Euclidean  plane.
The properties of hyperbola on sphere and complex cylinder are
described in~\cite{Fenchel989,Iversen1992}. Note that
\textit{Mathematica} computes the area hyperbolic tangent of real
or complex argument on a complex closed
cylinder~\cite{Fenchel989}.}
$\text{tanh}(x,y)=y/x=\sinh\varphi_h/\cosh\varphi_h=\tanh\varphi_h$.
As follows from Fig.~\ref{fig:circeHyperbola}b the range of the
hyperbolic tangent is $(-1...1)$ when
${\varphi_h=-\infty...\infty}$. Then, in the quadrants
\textit{1,2,3,4} we have, respectively, $\text{tanh}(+|x|,+|y|)$,
$\text{tanh}(-|x|,+|y|)$, $\text{tanh}(-|x|,-|y|)$, and
$\text{tanh}(+|x|,-|y|)$.

We shall assume that in GA the defining equation of the logarithm
is $\log\m{B}=\m{A}$, which takes into account only the principal
value (principal logarithm). To include multiple values we add a
free MV $\m{F}$,
\begin{equation}\label{logCL01A}
\log\m{B}=\m{A}+\m{F},\qquad\m{A},\m{B},\m{F}\in\cl{0}{1},
\end{equation}
that satisfies $\ee^\m{F}=1$. Equation~\eqref{logCL01A} is more
general because, as we shall see, it allows to include the
multiplicity into  GA logarithm  in case of higher ($n=3$)
dimensional GAs~\cite{AcusDargys2022b}. Let's apply the described
approach to Eq.~\eqref{LogCL01}
\begin{equation}\label{logCL01def}
\log\m{B}=\text{log}\magnitude{\m{B}}+\e{1}\text{arctan}(b_0,b_1)
\equiv \log r+\e{1}\varphi,
\end{equation}
where
$\magnitude{\m{B}}=\sqrt{\m{B}\gradeinverse{\m{B}}}=\sqrt{b_0^2+b_1^2}=r$
is the  radius $r$  (magnitude or norm of $\m{B}$)  and $\varphi$
is the angle between the horizontal axis and line that connects
the coordinate center with the point~$b$,
Fig.~\ref{fig:circeHyperbola}a. To include multiplicity in the
angle, a free term $\m{F}$ is added, $\log\m{B}=\m{A}+\m{F}$.
After substitution of $\m{F}=f_0+\e{1}f_1$ into $\ee^\m{F}=1$ and
using the trigonometric expansion similar to Eq.~\eqref{expCL01},
we find $\ee^{f_0}(\cos f_1+\e{1}\sin f_1)=1$, the solution of
which is $f_0=0$ and $f_1=2\pi k$, where $k\in\bbZ$. Thus, the
full solution in agreement with the complex function theory can be
written
\begin{equation}\label{LogCL01b}
\cl{0}{1}:\quad\log\m{B}=\text{log}\magnitude{\m{B}}+\e{1}(\varphi+2\pi
k),\quad 0\le\varphi<2\pi,\quad k\in\bbZ.
\end{equation}
At a fixed $r=\magnitude{\m{B}}$ this equation represents the
spiral with period $2\pi$ since  the argument ($0\le\varphi<
2\pi$)  increases by $2\pi$ after every single winding in the
``complex'' plane $\{b_0,b_1\}$.  The logarithm~ \eqref{LogCL01}
exists for all values of~$\m{B}$. Often it is assumes that the
principal logarithm is in the range $-\pi<\varphi<\pi$, then the
logarithm is
\begin{equation}\label{logCL01Mathematica}
\cl{0}{1}:\quad\log\m{B}=
\begin{cases}
 \text{log}\magnitude{\m{B}}+\e{1}\varphi&\quad\text{if\ }b_0>0\text{\ and\ } b_1\ne 0,\\
 \text{log}\magnitude{\m{B}}+\e{1}(\varphi+\pi)&\quad\text{if\ }b_0<0\text{\ and\ } b_1>0,\\
 \text{log}\magnitude{\m{B}}+\e{1}(\varphi-\pi)&\quad\text{if\ }b_0<0\text{\ and\ } b_1<0, \\
 \hline
b_0&\quad\text{if\ }b_0>0\text{\ and\ } b_1=0,\\
b_0+\e{1}\pi&\quad\text{if\ }b_0<0\text{\ and\ } b_1=0,\\
b_1+\e{1}\pi/2&\quad\text{if\ }b_0=0\text{\ and\ } b_1>0,\\
b_1-\e{1}\pi/2&\quad\text{if\ }b_0=0\text{\ and\ } b_1<0.\\
\end{cases}
\end{equation}
The first three expressions are the main formulas.  The remaining
represent special cases: they  show the behavior of logarithm on
the real and imaginary axis. When $b_0=b_1=0$ the logarithm is
undefined. The definition given by Eqs~\eqref{logCL01Mathematica}
and visualized in Fig.~\ref{fig:PrincipalValue}b frequently is met
in applications. It has been  accepted in ISO standards such as C
programming language and \textit{Mathematica}.

\subsection{Exponential and logarithm  of MV in \cl{1}{0}}
\begin{figure}
\centering
 \begin{minipage}[c]{0.45\textwidth}
  \centering
  \includegraphics[width=3.5cm]{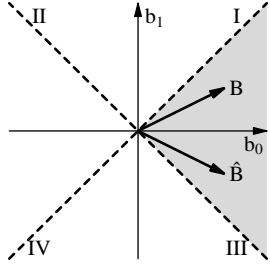}
 \end{minipage}%
  \begin{minipage}[c]{0.45\textwidth}
  \centering
\caption{\label{hyperplane}Hyperbolic plane $\{b_0,b_1\}$ that
represents \cl{1}{0} algebra. The arrows show MV
$\m{B}=b_0+\e{1}b_1$ and its conjugate~$\gradeinverse{\m{B}}$. The
dashed lines are asymptotes $b_0^2-b_1^2=0$. The principal square
root and logarithm exist in the  shaded sector where $b_0>b_1$.
I-IV are the hyperbolic plane quadrants.\label{hyperbolicPlane}}
  \end{minipage}
\end{figure}

For 1D algebras the inverse of MV $\m{B}=b_0+b_1\e{1}$ is
 \begin{equation}\label{inverse1D}
 \cl{0}{1}:\
 \m{B}^{-1}=\frac{\gradeinverse{\m{B}}}{\m{B}\gradeinverse{\m{B}}}=\frac{b_0-\e{1}b_1}{b_0^2+b_1^2};
\qquad
 \cl{1}{0}:\ \m{B}^{-1}=\frac{\gradeinverse{\m{B}}}{\m{B}\gradeinverse{\m{B}}}=\frac{b_0-\e{1}b_1}{b_0^2-b_1^2},
\end{equation}
and satisfies $\m{B}^{-1}\m{B}=\m{B}\m{B}^{-1}=1$. From
\eqref{inverse1D} follows that, in contrast to complex algebra
where each nonzero complex number has its inverse, in \cl{1}{0}
nonzero divisors appear  if $b_0^2=b_1^2$ as shown by dashed lines
in a hyperbolic plane in Fig.~\ref{hyperbolicPlane}. Since
$\e{1}^2>0$, \cl{1}{0} exponential may be expanded in hyperbolic
sine and cosine functions~\cite{Lounesto97},
\begin{equation}\label{expBhyperbolic}
\cl{1}{0}:\quad\ee^{\m{B}}=\ee^{b_0+b_1\e{1}}=\ee^{b_0}\ee^{b_1\e{1}}=\ee^{b_0}
\big(\cosh b_1+\e{1}\sinh b_1\big),
\end{equation}
where $b_0,b_1\in\bbR$. The  hyperbolic functions are monotonic
therefore the exponential in \cl{1}{0} inherits this property as
well.
In \cl{1}{0} a logarithm defining equation is $\log\m{B}=\m{A}$,
in  solution of which the hyperbolic functions and the identity
$\cosh^2 x-\sinh^2 x=1$ are to be used.  The following expression
for principal logarithm (the first formula) and special case (the
second formula ) is found:
\begin{equation}\label{logCL10a}
\cl{1}{0}:\quad\log\m{B}=
  \begin{cases}\log\sqrt{b_0^2-b_1^2}+\e{1}\text{artanh}(b_1/b_0);\ b_0>0\ \text{and}\ b_0^2>b_1^2,\\
    \frac{1}{2}(\log(0_+{}) + \log (2 b_0))\pm \e{1}\frac{1}{2}(-\log(0_{+}) +  \log (2
    b_0);\  b_1=b_0,
  \end{cases}
\end{equation}
where $\text{artanh}$ is the area tangent function,
$-1<\text{artanh}(b_1/b_0)<1$. The scalar part
$\log\sqrt{b_0^2-b_1^2}$ exists if $b_0^2>b_1^2$. The logarithm
has a genuine value if a pair $\{b_0,b_1\}$ is in the shaded
sector of Fig.~\ref{hyperbolicPlane}. Thus, the existence of both
the logarithm and the square root are determined by condition
$b_0^2>b_2^2$. The special case belongs to asymptotes $b_0=b_1$,
where $\log(0_{+})$ is the logarithm of a point infinitesimally
close to zero. This term vanishes in $\exp{(\log{\m{B}})}=\m{B}$
(see Example~2). The first equation of~\eqref{logCL10a} can be
rewritten in hyperbola parameters in
Fig.~\ref{fig:circeHyperbola}b. Since
$\m{B}=r(\cosh\varphi_h+\e{1}\sinh\varphi_h)$, where $r=a_0$ is
the radius ($r=1$ in Fig.~\ref{fig:circeHyperbola}b) we have
$b_0=r\cosh\varphi_h$ and $b_1=r\sinh\varphi_h$. Since
$r=\sqrt{b_0^2-b_1^2}$ and $\tanh\varphi_h=b_1/b_0$, we have that
$\log\m{B}=\log r+\e{1}\varphi_h$ which is to be compared with
Eq.~\eqref{LogCL01}.

To find the free term $\m{F}$  we solve $\exp({\m{F}})=1$, for
this purpose bringing into play Eq.~\eqref{expBhyperbolic},
\begin{equation}
\ee^{\m{F}}=\ee^{f_0+\e{1}f_1}=\ee^{f_0}\big(\cosh
\magnitude{f_1}+\e{1}\sinh{\magnitude{f_1}}\big)=1.
\end{equation}
This equation can be satisfied if $f_0=f_1=0$. So, in this algebra
we have only the principal logarithm.

\vspace{3mm} \textbf{Example~1.} \cl{1}{0}:\quad If $b_0^2>b_1^2$,
$b_0>0$, and  $\m{B}=3\pm2\e{1}$  then
$\log\m{B}=\log\sqrt{5}\pm\e{1}\text{artanh}\frac{2}{3}=a_0\pm\e{1}a_1$.
The exponential of logarithm gives $\exp(\log\m{B})=\m{B}$. If
$b_0^2>b_1^2$, $b_0<0$ and $\m{B}^{\prime}=-3\pm2\e{1}$, then
$\log\m{B}^{\prime}=\log\sqrt{5}\pm\e{1}\text{artanh}\frac{2}{3}=a_0\pm\e{1}a_1\,$.
The answer is wrong since the initial MV is returned with an
opposite sign: $\exp(\log\m{B}^{\prime})=-\m{B}^{\prime}$.

 \vspace{2mm}
\textbf{Example~2.}  \cl{1}{0}:\quad $b_0=b_1$,  $\m{B}=2+
2\e{1}$. In this case the second formula of~\eqref{logCL10a}
should be used:
 $\log\m{B}=\frac12(\log4+\log 0_{+})+\frac12\e{1}(\log4-\log 0_{+})$. Then
 $\exp(\log\m{B})=\frac12\big(4(1+\e{1})\big)+\frac12\big(\ee^{\log0_{+}}(-1+\e{1})\big)
 \to 2(1+\e{1})$ which in the limit $\log0_{+}\to -\infty$ gives  $\m{B}=2+
2\e{1}$ that represents a point on the asymptote.

\section{\label{sec:ExpLog2D} Exponential and logarithm
in 2D algebras}

\subsection{\label{qvector}Quaternionic `vector'}

The following defining equations for  exponential and logarithm in
2D GAs are used,  $\exp\m{B}=\m{A}$ and $\log\m{B}=\m{A}$, where
MVs $\m{A}$ and $\m{B}$ belong to the same algebra. It is
convenient to introduce base-free MVs
$A^{\prime}=\ba+b_{12}\e{12}$ and $B^{\prime}=\bb+b_{12}\e{12}$,
where $\ba$ and $\bb$ are vectors in $\e{12}$ plane. In analogy to
Hamilton quaternion theory~\cite{Gurlebeck1997,Morais14}, in the
following we shall treat the quantity $A^{\prime}$ as a 3D
`vector'. Introduction of such a `vector' appears very helpful in
calculating the exponential as well as logarithm in all 2D
algebras. Thus, a full MV in 2D algebras may be represented as a
sum of scalar and `vector':
\begin{equation}
\m{A}=a_0+A^{\prime},\quad
A^{\prime}=a_1\e{1}+a_2\e{2}+a_{12}\e{12}.
\end{equation}
In \cl{0}{2} the squares of all three basis elements  satisfy
$\e{1}^2=\e{2}^2=\e{12}^2=-1$ and $\e{1}\e{2}\e{12}=-1$.
Similarly, in the Hamilton quaternion
algebra~\cite{Gurlebeck1997,Morais14} a set of three imaginary
units $\{\ii,\jj,\kk\}$ satisfy $\ii^2=\jj^2=\kk^2=-1$ and
$\ii\jj\kk=-1$. In 3D Euclidean space the quaternionic vector is
defined by $\bv=a_1\ii+a_2\jj+a_{12}\kk$, the square of which is a
negative number. The same property is satisfied by `vector',
$(A^{\prime})^2\equiv A^{\prime 2}=-a_1^2-a_2^2-a_{12}^2<0$, where
$a_1,\ a_2$ and $a_{12}$ are the real numbers. Thus, the MV
$A^{\prime}$ is equivalent to quaternion vector and  $A^{\prime}$
may be treated exactly in the same way  as the  Hamilton vector
$\bv=a_1\ii+a_2\jj+a_{12}\kk$. For \cl{0}{2} 'vector' $A^{\prime}$
the norm is defined by $\magnitude{A^\prime}=\sqrt{-A^{\prime
2}}$.

Because \cl{1}{1} and \cl{2}{0} are not division algebras, i.e. in
these algebras not every MV has inverse, for these algebras we
have different cases.  Now  $A^{\prime 2}\equiv (A^\prime)^2$ may
be either positive or negative, or even zero.  The first
(positive) case, as we shall see, is related to hyperbolic
functions, while the second is related to trigonometric functions.
Both cases will be investigated separately in \cl{1}{1} and
\cl{2}{0} algebras. The  semi-norm of 'vector' $A^{\prime}$ is
defined by $\lVert A^\prime\rVert=\sqrt{\abs(A^{\prime 2})}\ge 0$.

\begin{center}
\begin{table}
\begin{tabular}{lll}\hline
 &  $\exp(\m{B})=\exp(b_0+B^{\prime})$\ \ \qquad\qquad\qquad  $ -B^{\prime}=b_1\e{1}+b_2\e{2}+b_{12}\e{12}$ &\\
\hline
  \cl{0}{2} & $\begin{cases}\ee^{b_0}\Big(\cos\magnitude{ B^{\prime}}+\frac{B^{\prime}}{\magnitude{ B^{\prime}}}\sin\magnitude{ B^{\prime}}\Big)& \qquad\quad\ B^{\prime 2}=b_1^2+b_2^2+b_{12}^2\,>0\\
   \ee^{b_0}&\qquad\quad\ B^{\prime}=0
   \end{cases}$ &  \\[10pt]
 \cl{1}{1} &
  $\begin{cases}
    \ee^{b_0}\Big(\cosh\lVert B^{\prime}\rVert+\frac{B^{\prime}}{\lVert B^{\prime}\rVert}\sinh\lVert B^{\prime}\rVert\Big) &\quad B^{\prime 2}=b_1^2-b_2^2+b_{12}^2\,> 0 \\
    \ee^{b_0}(1+B^{\prime}) &\quad B^{\prime 2}=b_1^2-b_2^2+b_{12}^2=0\\
    \ee^{b_0}\Big(\cos\lVert B^{\prime}\rVert+\frac{B^{\prime}}{\lVert B^{\prime}\rVert }\sin\lVert B^{\prime}\rVert\Big) &
   \quad B^{\prime 2} =b_1^2-b_2^2+b_{12}^2\,< 0
  \end{cases}$ &
\\[10pt]
\cl{2}{0} &
  $\begin{cases}
    \ee^{b_0}\Big(\cosh\lVert B^{\prime}\rVert+\frac{B^{\prime}}{\lVert B^{\prime}\rVert}\sinh\lVert B^{\prime}\rVert\Big)&\quad B^{\prime 2}=b_1^2+b_2^2-b_{12}^2\,> 0 \\
     \ee^{b_0}(1+B^{\prime}) &\quad B^{\prime 2}=b_1^2+b_2^2-b_{12}^2=0  \\
    \ee^{b_0}\Big(\cos\lVert B^{\prime}\rVert+\frac{B^{\prime}}{\lVert B^{\prime}\rVert}\sin\lVert B^{\prime}\rVert\Big) &
    \quad B^{\prime2}=b_1^2+b_2^2-b_{12}^2\,< 0
  \end{cases}$ & \\ \hline
\end{tabular}
 \caption{\label{exp2D}
Exponentials  of general MV
$\m{B}=b_0+B^{\prime}=b_0+b_1\e{1}+b_2\e{2}+b_{12}\e{12}$ in 2D
  GAs. $\magnitude{B^\prime}=\sqrt{-B^{\prime 2}}$  and  $\lVert B^{\prime}\rVert=\sqrt{\abs(B^{\prime 2})}$ are
real numbers that represent the norm and semi-norm, respectively.
The trigonometric functions appear when $B^{\prime 2}<0$ while
hyperbolic  when  $B^{\prime 2}>0$. Zero values of the semi-norm
corresponds to $\lim_{x\to 0}\tfrac{\sin(x)}{x}=\lim_{x\to
  0}\tfrac{\sinh(x)}{x}=1$.
 }
\end{table}
\end{center}
 \subsection{\label{logarithm2D} Exponentials of MV in 2D algebras}
In Table~\ref{exp2D}, two-dimensional exponentials in expanded
form including the case of null MV (when $\m{B}^2=0$) are
summarized. The structure of the formulas  reminds de
Moivre's-Euler's rules. For \cl{0}{2} only trigonometric functions
appear. For algebras \cl{1}{1} and \cl{2}{0}  also hyperbolic
functions appear if $B^{\prime 2}>0$. In case of \cl{0}{2} which
is isomorphic to Hamilton quaternion the exponential  formula can
be found easily if the property
$\bigl(B^\prime/\magnitude{B^\prime}\bigr)^2=-1$ is taken into
account. Since $B^\prime/\magnitude{B^\prime}$ behaves like an
imaginary unit we can write at once
\begin{equation}
\ee^{B^\prime}= \cos\magnitude{B^\prime}+
\frac{B^\prime}{\magnitude{ B^\prime}}\sin\magnitude{B^\prime}.
\end{equation} Then, the exponential of $\m{B}=b_0+B^\prime$ is
\begin{equation}
\cl{0}{2}:\
\ee^{\m{B}}=\ee^{b_0+B^\prime}=\ee^{b_0}\ee^{B^\prime}=\ee^{b_0}\bigl(\cos\magnitude{
B^\prime}+ \frac{B^\prime}{\magnitude{ B^\prime}}\sin\magnitude{
B^\prime}\bigr).
\end{equation}

In the remaining algebras the  different normalization must be
used. The square of a normalized `vector' now is
$\bigl(B^\prime/\lVert B^\prime\rVert \bigr)^2=\pm1$ and apart
from trigonometric, in addition, hyperbolic functions for plus
sign appear,
\begin{equation}\label{expCL11CL30}
\cl{1}{1},\cl{2}{0}:\ \ee^{\m{B}}=\ee^{b_0}\ee^{B^\prime}=
\begin{cases}\ee^{b_0}\bigl(\cos\lVert
B^\prime\rVert+ \frac{B^\prime}{\lVert B^\prime\rVert}\sin\lVert
B^\prime\rVert\bigr),\quad B^{\prime 2}<0,\\
\ee^{b_0}\bigl(\cosh\lVert B^\prime\rVert+ \frac{B^\prime}{\lVert
B^\prime\rVert}\sinh\lVert B^\prime\rVert\bigr),\quad B^{\prime
2}>0.
 \end{cases}
 \end{equation}
Thus, in \cl{1}{1} and \cl{2}{0} algebras  depending on sign of
$B^{\prime 2}$ and coefficient values in the semi-norm, the
exponentials may be expanded either in trigonometric or in
hyperbolic functions and as a result mat be periodic  or
monotonic. Finally, in Table~\ref{exp2D} the exponential
$\ee^{\m{B}}=\ee^{b_0}\big(1+B^\prime\big)$ comes from the null MV
the square of which  nullifies  $B^{\prime 2}=0$ and yields a
linearly dependence on $B^{\prime}$. Recently we have
found~\cite{AcusDargys2022a} that in three-dimensional GAs (and
probably in higher dimensional spaces) the entanglement or mixing
of vector and bivector components may tale place, so that in the
expanded form the exponential loses de Moivre's-Euler's formula
structure. The latter is regained if both the vector and bivector
lie in the same plane. This is in agreement with the present 2D
formulas where the vector and bivector  are always in
$\e{1}-\e{2}$ plane.

\vspace{2mm}
\textbf{Example~3.}  $\m{B}=2+5\e{1}-4\e{2}-7\e{12}=2+B^{\prime}$.\\
 \cl{0}{2}:\
 $\exp\m{B}=\ee^2\left(\cos\sqrt{90}+\frac{5\e{1}-4\e{2}-7\e{12}}{\sqrt{90}}\sin\sqrt{90}\right), B^{\prime2}=-90,\magnitude{B^\prime}=\sqrt{90}$,\\
\cl{1}{1}:\
$\exp\m{B}=\ee^2\left(\cosh\sqrt{58}+\frac{5\e{1}-4\e{2}-7\e{12}}{\sqrt{58}}\sinh\sqrt{58}\right),\ B^{\prime2}=58,\lVert B^\prime\lVert=\sqrt{58}$,\\
\cl{2}{0}:\
$\exp\m{B}=\ee^2\left(\cos\sqrt{8}+\frac{5\e{1}-4\e{2}-7\e{12}}{\sqrt{8}}\sin\sqrt{8}\right),\quad B^{\prime2}=-8,\lVert B^\prime\rVert=\sqrt{8}$.

\subsection{Products of exponentials} \label{sec:expProducs}
Using  Table~\ref{exp2D} it is easy to calculate the geometric
product of two exponentials. For example, for trigonometric
functions in \cl{0}{2}, when $B^{\prime2}<0$, we find
\begin{equation}\label{prod2exp}\begin{split}
&\ee^{\m{A}}\ee^{\m{B}}=\ee^{a_0+A^\prime}\ee^{b_0+B^\prime}=\ee^{a_0+b_0}\left(\cos\magnitude{A^\prime}\cos\magnitude{B^\prime}
+\frac{\langle A^\prime B^\prime\rangle_0}{\magnitude{A^\prime}\magnitude{B^\prime}}\sin\magnitude{A^\prime}\sin\magnitude{B^\prime}\right)+\\
&\ee^{a_0+b_0}\left(\frac{A^\prime}{\magnitude{A^\prime}}\sin\magnitude{A^\prime}\cos\magnitude{B^\prime}+
\frac{B^\prime}{\magnitude{B^\prime}}\sin\magnitude{B^\prime}\cos\magnitude{A^\prime}+\frac{1}{2}
\frac{[A^\prime,B^\prime]}{\magnitude{A^\prime}\magnitude{B^\prime}}\sin\magnitude{A^\prime}\sin\magnitude{B^\prime}\right).
\end{split}\end{equation}
When $A^\prime=B^\prime$ the commutator $[A^\prime,B^\prime]=0$
and the Eq~\eqref{prod2exp} reduces to double $2\m{A}$ argument
exponential. For remaining algebras the norm should be replaced by
semi-norm. Below, particular cases follow from~\eqref{prod2exp}.

\textit{Case~1. Product of vectorial exponentials}. If $\m{A}$ and
$\m{B}$ represent vectors $\ba=a_{1}\e{1}+a_{2}\e{2}$, and
$\bb=b_{1}\e{1}+b_{2}\e{2}$, then
\begin{equation}\begin{split}
\ee^{\ba}\ee^{\bb}=&\cos|\ba|\cos|\bb|-\cos\theta\sin|\ba|\sin|\bb|+\\
&\frac{\ba}{|\ba|}\sin|\ba|\cos|\bb|+\frac{\bb}{|\bb|}\sin|\bb|\cos|\ba|+
\e{12}\sin\theta\sin|\ba|\sin|\bb|,
\end{split}\end{equation} where $\theta$ is the angle between vectors $\ba$ and
$\bb$.

 \textit{Case~2. Product of bivectorial
exponentials}. If $\m{A}$ and $\m{B}$ are simple bivectors
$\cA=a_{12}\e{12}$, $\cB=b_{12}\e{12}$, then
\begin{equation}
\ee^{\cA}\ee^{\cB}=\ee^{\cA+\cB}=
\cos\magnitude{\cA+\cB}+\e{12}\sin\magnitude{\cA+\cB},
\end{equation}
where $\magnitude{\cA+\cB}=\sqrt{(\cA+\cB)\reverse{(\cA+\cB)}}$.
Since $\cA+\cB$ is the bivector and $(\cA+\cB)^2<0$ this formula
follows  directly.

\textit{Case~3. Product of vector and bivector exponentials}:
\begin{equation}
\ee^{\ba}\ee^{\cB}=(\cos{|\ba|}+\frac{\ba}{|\ba|}\sin{|\ba}|)(\cos|\cB|+\e{12}\sin|\cB|)
\end{equation}

 \textit{Case~4}. The commutator also vanishes  if the
coefficients satisfy: $a_{1}b_{12}=b_{1}a_{12}$,
$a_{2}b_{12}=b_{2}a_{12}$ and $a_{2}b_{1}=b_{2}a_{1}$.  Since in
this case  $[\m{A},\m{B}]=[A^\prime,B^\prime]$  we have
$\ee^{\m{A}}\ee^{\m{B}}=\ee^{\m{A}+\m{B}}$ and
Eq.~\eqref{prod2exp} reduced to
\begin{equation}
\ee^{\m{A}}\ee^{\m{B}}=\ee^{\m{A}+\m{B}}=\ee^{\m{B}}\ee^{\m{A}}=
\ee^{a_0+b_0}\Big(\cos\magnitude{A^\prime+B^\prime}+\frac{A^\prime+B^\prime}{\magnitude{A^\prime+B^\prime}}\sin\magnitude{A^\prime+B^\prime}\Big).
\end{equation}
When $B^{^\prime2}>0$, similar formulas exist for hyperbolic
functions.

\section{Logarithm of MV in 2D algebras}\label{sec:Log3D}

The approach to commutative algebras in Subsec.~\ref{logCL01} here
is generalized to 2D algebras. The 'vector' property
$B^{\prime2}\gtreqqless0$ allows to get 2 dimensional logarithm
formulas that are very similar to those found in  1D case but with
basis vector  $\e{1}$ replaced by unit multivector
$B^\prime/\magnitude{B^\prime}$ or $B^\prime/\lVert
B^\prime\rVert$.

\subsection{\label{algebrasCL02} \cl{0}{2} algebra}

In this algebra according to Table~\ref{exp2D} the norm (or
magnitude) of $B^\prime$ is $\magnitude{B^\prime}=\sqrt{-B^{\prime
2}}=\sqrt{B^\prime\cliffordconjugate{B^\prime}}
=\sqrt{b_1^2+b_2^2+b_{12}^2}$.  The logarithm defining equation is
$\log\m{B}=\m{A}$, where $\m{B}$ is a given MV, and coefficients
of~$\m{A}$ are to be determined. Since $B^{\prime2}<0$ the
exponential of logarithm can be expanded by  trigonometric
functions,
\begin{equation}
\ee^{\log\m{B}}=\ee^{\m{A}}=\ee^{a_0+A^\prime}=\ee^{a_0}\big(\cos\magnitude{
A^\prime}+\frac{A^\prime}{\magnitude{A^\prime}}\sin\magnitude{A^\prime}\big),
\end{equation}
from which  we write the following relation between 'vectors'
$B^\prime$ and $A^\prime$,
\begin{equation}\label{b0B}
b_0+B^\prime=\ee^{a_0}\big(\cos\magnitude{A^\prime}+\frac{A^\prime}
{\magnitude{A^\prime}}\sin\magnitude{A^\prime}\big).
\end{equation}
Equation~\eqref{b0B} can be rewritten as a system of two
equations,
\begin{equation}\label{equations1}
b_0=\ee^{a_0}\cos\magnitude{A^\prime},\quad
B^\prime=\ee^{a_0}\frac{A^\prime}{\magnitude{A^\prime}}\sin\magnitude{A^\prime},
\end{equation}
where the second equation, in fact, represents three scalar
equations. The system~\eqref{equations1}  can be solved with
respect to $a_0$ and $A^\prime$ in the following way. After
squaring both sides of~\eqref{equations1} and noting that in
\cl{0}{2} $B^{\prime2}=-\magnitude{B^\prime}^2$, we have
\begin{equation}\label{equations3a}
b_0^2=\ee^{2a_0}\cos^2\magnitude{A^\prime},\quad
\magnitude{B^\prime}^2=\ee^{2a_0}\sin^2\magnitude{A^\prime}.
\end{equation}
The sum  gives   $b_0^2+\magnitude{B^\prime}^2 =\ee^{2a_0}$ from
which and $b_0^2+\magnitude{B^\prime}^2=\magnitude{ \m{B}}^2$
follows
\begin{equation}\label{a0B}
a_0=\log\magnitude{\m{B}}.
\end{equation}
The ratio of equations in~\eqref{equations3a} gives
$\magnitude{B^\prime}/b_0=\tan\magnitude{A^\prime}$.The inverse of
the latter is
\begin{equation}\label{magnitudeB}
\magnitude{A^\prime}=\text{arctan}\big(\magnitude{
B^\prime}/b_0\big).
\end{equation}
To express the `vector' $A^\prime$ in terms of $B^\prime$, the
second equation in~\eqref{equations1} is divided by the first,
\begin{equation}\label{Bb0}
\frac{B^\prime}{b_0}=\frac{A^\prime}{\magnitude{A^\prime}}\tan\magnitude{
A^\prime}.
\end{equation}
As follows from!\eqref{magnitudeB}
$\tan\magnitude{A^\prime}=\magnitude{B^\prime}/b_0$, therefore,
the Eq.~\eqref{Bb0} reduces to
\begin{equation}\label{Aa0}
\frac{B^\prime}{b_0}=\frac{A^\prime}{\magnitude{
A^\prime}}\frac{\magnitude{B^\prime}}{b_0},
\end{equation}
from which the property $B^\prime/\magnitude{
B^\prime}=A^\prime/\magnitude{A^\prime}$, i.e. $B^\prime$ and
$A^\prime$ are parallel in $\{\ii,\jj,\kk\}$ space, follows. The
latter along with \eqref{a0B} allow to get
\begin{equation}
\log\m{B}=a_0+A^\prime=\log\magnitude{\m{B}}+\magnitude{
A^\prime}\frac{B^\prime}{\magnitude{B^\prime}}.
\end{equation}
Finally, the needed generic logarithm formula is
\begin{equation}\label{logBinCL02}
\cl{0}{2}:\ \log\m{B}=
\log\magnitude{\m{B}}+\frac{B^\prime}{\magnitude{
B^\prime}}\left(\text{arctan}\frac{\magnitude{
B^\prime}}{b_0}\right),\quad B^{\prime2}<0.
\end{equation}

To logarithm ~\eqref{logBinCL02} we may add a free MV $\m{F}=2\pi
k \hat{F}^\prime$, where $\hat{F}^\prime$ plays the role of
imaginary unit, $(\hat{F}^\prime)^2=-1$. In addition, it satisfies
$\exp(2\pi k \hat{F}^\prime)=1$ and
$\magnitude{\hat{F}^\prime}=1$. As we shall see the free MV takes
into account the multi-valuedness of arc tangent. Then,
$\text{log}{\m{B}}=\m{A}+\m{F}=\m{A}+2\pi k \hat{F}^\prime$. Since
$\m{F}=f_0+f_1\e{1}+f_2\e{2}+f_{12}\e{12}=f_0+F^\prime$ and
$\magnitude{
F^\prime}=\sqrt{-F^{\prime2}}=\big(F^\prime\cliffordconjugate{F^\prime}\big)^{1/2}=\sqrt{f_1^2+f_2^2+f_{12}^2}$,
we have
\begin{equation}
\ee^{\m{F}}=\ee^{f_0}\big(\cos\magnitude{
F^\prime}+\frac{F^\prime}{\magnitude{F^\prime}}\sin\magnitude{
F^\prime}\big)=1,
\end{equation}
which is satisfied if $f_0=0$ and
$\magnitude{F^{\prime}}=\sqrt{f_1^2+f_2^2+f_{12}^2}=2\pi k,\quad
k\in\bbZ.$ Thus, in \cl{0}{2}  we have that the free MV is
$\m{F}=2\pi k \hat{F}^\prime=2\pi
k\frac{f_1\e{1}+f_2\e{2}+f_{12}\e{12}}{\sqrt{f_1^2+f_2^2+f_{12}^2}}=2\pi
k\tfrac{F^\prime}{\magnitude{F^\prime}}$ that represents all
possible `vectors' the ends of which lie on a sphere of radius
equal to~$1$ in the 3D anti-Euclidean space $\{\ii,\jj,\kk\}$.
When $k=0$ then $\magnitude{F^\prime}=0$ and $\m{F}=0$, because
$f_0=0$. Thus, we conclude that the generic solution of equation
$\text{log}{\m{B}}=\m{A}$ represents the principal value of
argument $\varphi=\arctan\big(\magnitude{B^\prime}/b_0\big)=
\arctan\big(\sqrt{b_1^2+b_2^2+b_3^2}/b_0\big)$ in the range $0\le
\varphi<2\pi$ if $k=0$. When  multi-valuedness  is included the
Hamilton quaternion logarithm takes the form
\begin{equation}\label{logC02all}
\log\m{B}+\m{F}=\log\magnitude{\m{B}}+ \frac{B^\prime}{\magnitude{
B^\prime}}\Big(\arctan\frac{\magnitude{B^\prime}}{b_0}+2\pi
k\Big),\quad B^{\prime2}<0,\quad k\in\bbZ.
\end{equation}
$\magnitude{\m{B}}=(b_0^2+\magnitude{B^\prime}^2)^{1/2}$. The
formula \eqref{logC02all} satisfies
$\ee^{\text{log}\m{B}+\m{F}}=\m{B}$ for all integers $k$. Thus, in
\cl{0}{2} the free MV is
$\m{F}=F^{\prime}=(B^\prime/\magnitude{B^\prime})2\pi k$, where
the `vector' $(B^\prime/\magnitude{B^\prime})$ plays the role of
an imaginary unit (compare with Eq.~\eqref{LogCL01}). When $k=0$
we return back to the principal logarithm. Thus, after replacement
of the arc tangent by a double-argument arc tangent in order to
take account of all four quadrants correctly, the generic
formula~\eqref{logC02all} with  special cases included becomes:
\begin{align}\label{logCL02}
\cl{0}{2}:\quad \log(\m{B})= &\begin{cases}
 \log\magnitude{\m{B}} +
 \Bigl(\arctan\bigl(b_0,\magnitude{B^\prime}\bigr)+2\pi k\Bigr)
 \frac{B^\prime}{\magnitude{B^\prime}},
 \quad\magnitude{B^\prime}\neq 0,\\
  \log(b_0)+2\pi k\,\hat{F}^{\prime},
  \quad (\magnitude{B^\prime}= 0)\land (b_0>0),\\
  \log(-b_0)+\pi(2k+1)\, \hat{F}^{\prime},
 \quad (\magnitude{B^\prime}= 0)\land (b_0 < 0).
\end{cases}
\end{align}
To summarize,  we have shown that, similar to complex number
logarithm,  the Hamilton number logarithm  is a multi-valued
function too.

\vspace{3mm}
\textbf{Example~4.} \cl{0}{2}: \quad
$\m{B}=2+4\e{1}-5\e{2}-\e{12}$.\quad
\quad$\magnitude{\m{B}}=\sqrt{46}$, $B^{\prime2}=-42<0$,
\quad$\magnitude{B^\prime}=\sqrt{42}$.  The principal logarithm is
$\log\m{B}=\log\sqrt{46}+\left(\text{arctan}\frac{\sqrt{42}}{2}\right)\frac{4\e{1}-5\e{2}-\e{12}}{\sqrt{42}}$
 $\approx 1.914+4.662\e{1}-5.826\e{2}-1.165\e{12}$\,.
After exponentiation $\ee^{\text{log}\m{B}}$ we recover the initial MV  $\m{B}$.\\

 \subsection{\cl{1}{1} and \cl{2}{0} algebras, when $B^{\prime 2}\le 0$}
When $B^{\prime 2}=-B^{\prime }\cliffordconjugate{B^{\prime}}<0$,
the exponentials for both \cl{1}{1} and \cl{2}{0} a;algebra are
expressed in trigonometric functions in the same way as for
\cl{0}{2} but the norm replaced by semi-norm (see
Table~\ref{exp2D}). The free MV $\m{F}=f_0+\hat{F}^\prime$ also
satisfies the condition $\ee^{2\pi k \m{F}}=1$ from which  we have
$f_0=0$ and
\begin{equation}\begin{split}\label{hatFCL11CL29}
&  \cl{1}{1}:\quad
\hat{F}^\prime=f_1\e{1}+f_{12}\e{12}+\sqrt{1+f_1^2+f_{12}^2}\
\e{2},\\
& \cl{2}{0}:\quad
\hat{F}^\prime=f_1\e{1}+f_2\e{2}-\sqrt{1+f_1^2+f_2^2}\ \e{12},
\end{split}\end{equation}
with properties $\lVert \hat{F}^\prime\rVert=1$ and $\hat{F}^{\prime 2}=-1$.

Generic logarithm for both \cl{1}{1} and \cl{2}{0} are given by
equations similar to ~\eqref{logCL02} but with $\hat{F}^\prime$
replaced by~\eqref{hatFCL11CL29} and norm replaced by semi-norm,
\begin{align}\label{logCL11CL20}
\cl{1}{1},\cl{2}{0}:\ \log(\m{B})= &\begin{cases}
  \log(\lVert\m{B}\rVert)+\bigl(\arctan (b_0,\lVert B^{\prime}\rVert)+2\pi k\bigr)\frac{B^{\prime}}{\lVert B^{\prime}\rVert} ,
  \quad (B^{\prime 2}<0),\\
  \log(b_0)+B^\prime/b_0,\quad ( B^{\prime 2} =0)\land (b_0>0),\\
  \log(-b_0)+\pi(2k+1)\, \hat{F}^{\prime},
 \quad (B^\prime= 0)\land (b_0 < 0).
\end{cases}
\end{align}
The last two equations represent special cases. Now  the logarithm
of MV exists when $\lVert B^\prime\rVert\ne 0$ and $\lVert
B^\prime\rVert=0$. More specific cases are presented in
Subsec.~\ref{specificCases}.

 \vspace{3mm}
\textbf{Example~5.}  \cl{1}{1}:\quad
$\m{B}=2+4\e{1}-5\e{2}-\e{12}$,\quad$B^{\prime2}=-8$,
 \quad$\lVert \m{B}\rVert=\sqrt{12}$,\quad$\lVert
B^\prime\rVert=\sqrt{8}$. The answer
$\log\m{B}=\tfrac12\log(12)+\tfrac{1}{2\sqrt{2}}\left(\text{arctan}\sqrt{2}+2\pi
k\right)(4\e{1}-5\e{2}-\e{12})$
 satisfies  $\ee^{\log\m{B}}=\m{B}$.\\

\textbf{Example~6.}  \cl{2}{0},\quad
$\m{B}=2+5\e{1}-4\e{2}-7\e{12}$\quad$B^{\prime2}=-8$.
 \quad$\lVert B^\prime\rVert=\sqrt{8}$\quad$\lVert
\m{B}\rVert=\sqrt{12}$.\quad Answer:
 $\log\m{B}=\frac{\log(12)}{2}+\tfrac{1}{2\sqrt{2}}(\text{arctan}(\sqrt{2})+2\pi k)(5\e{1}-4\e{2}-7\e{12})$.
The answer satisfies  $\ee^{\log\m{B}}=\m{B}$.\\

\textbf{Example 7.} \cl{2}{0}. $\m{B}=2+3 \e{1}-4 \e{2}-5 \e{12}$;
$B^{\prime 2}=0$, $\lVert B^{\prime}\rVert=0$, $\lVert
\m{B}\rVert=2$.
\\Answer: $\log(\m{B})=\log(b_0)+B^\prime/b_0=\log(2)+\frac12 (3
\e{1}-4 \e{2}-5 \e{12})$.\ $\ee^{\log\m{B}}=\m{B}$..

\vspace{1mm}
 \textbf{Example~8.} \cl{2}{0}.\quad
$\m{B}=-2$,$B^{\prime2}=0$, $\lVert B^\prime\rVert=0$.\\
$\log\m{B}=\log(-b_0)+\pi(2k+1)\hat{F}^\prime=\log
2+\pi(2k+1)(f_1\e{1}+f_2\e{2}-\sqrt{1+f_1^2+f_2^2})\e{12}$.
The answer satisfies  $\ee^{\log\m{B}}=-2$.\\

\subsection{\cl{1}{1} and \cl{2}{0} algebras, when  $B^{\prime 2}>0$}
When  $B^{\prime 2}>0$, calculations proceed in a similar way.
Therefore, only intermediate results are put down briefly. Let the
general MV be $\m{B}=b_0+B^\prime$. The semi-norm in \cl{1}{1} is
$\lVert B^\prime\rVert=\sqrt{b_1^2-b_2^2+b_{12}^2}$ and in
\cl{2}{0} it  is $\lVert
B^\prime\rVert=\sqrt{b_1^2+b_2^2-b_{12}^2}0$, where expression
under the root should be positive. As in previous case the
defining equation is $\log{\m{B}}=\m{A}$, where
$\m{A}=a_0+A^\prime$. Since the square of $A^\prime$ now is
positive scalar, $A^{^\prime2}>0$, in agreement with the
Table~\ref{exp2D}, the exponential is expanded in hyperbolic
functions,
\begin{equation}
\ee^{\log\m{B}}=\ee^{a_0+A^\prime}=\ee^{a_0}\big(\cosh\lVert
A^\prime\rVert+\frac{A^\prime}{\lVert A^\prime\rVert}\sinh\lVert
A^\prime\rVert\big).
\end{equation}
Thus, we have the following relation between `vectors' $B^\prime$
and $A^\prime$,
\begin{equation}
b_0+B^\prime=\ee^{a_0}\big(\cosh\lVert
A^\prime\rVert+\frac{A^\prime}{\lVert A^\prime\rVert}\sinh\lVert
A^\prime\rVert\big),
\end{equation}
that may be rewritten as a system of equations
\begin{equation}\label{equations5b}
b_0=\ee^{a_0}\cosh\lVert A^\prime\rVert,\quad
B^\prime=\ee^{a_0}\frac{A^\prime}{\lVert
A^\prime\rVert}\sinh\lVert A^\prime\rVert.
\end{equation}
Squaring of Eqs.~\eqref{equations5b} and the property
$B^{\prime2}=\lVert B^\prime\rVert^2$  gives
\begin{equation}\label{equations7}
b_0^2=\ee^{2a_0}\cosh^2\lVert A^\prime\rVert,\quad \lVert
B^\prime\rVert^2=\ee^{2a_0}\sinh^2\lVert A^\prime\rVert.
\end{equation}
Now, applying  the property
$\cosh^2{A^\prime}-\sinh^2{A^\prime}=1$,  the difference of
equations in~\eqref{equations7} yields the scalar equation
$b_0^2-\lVert B^\prime\rVert^2=\ee^{2a_0}$, from which and the
relation $\lVert \m{B}\rVert^2=b_0^2-\lVert B^\prime\rVert^2$,
follows $a_0=\log\lVert \m{B}\rVert$. Also, from
Eq~\eqref{equations7} we have the ratio $\lVert
B^\prime\rVert/b_0=\tanh\lVert A^\prime\rVert$, from which we find
equation analogous  Eq~\eqref{magnitudeB}, $\lVert
{A^\prime\lVert=\text{artanh}\big(\lVert
B^\prime}\rVert/b_0\big)$, where $\text{artanh}$ is the area
hyperbolic tangent. To express~$A^\prime$ in terms of $B^\prime$,
we divide equations in~\eqref{equations5b},
\begin{equation}\label{Aa02}
\frac{B^\prime}{b_0}=\frac{A^\prime}{\lVert
A^\prime\rVert}\tanh\lVert A^\prime\rVert.
\end{equation}
Since $\tanh\lVert A^\prime\rVert=\lVert B^\prime\rVert/b_0$, the
Eq.~\eqref{Aa02} can be reduced to
\begin{equation}\label{Aa03}
\frac{B^\prime}{b_0}=\frac{A^\prime}{\lVert
A^\prime\rVert}\frac{\magnitude{B^\prime}}{b_0}.
\end{equation}
from which the property $B^\prime/\lVert
B^\prime\rVert=A^\prime/\lVert A^\prime\rVert$ follows. The latter
allows to get the required formula for the principal logarithm,
\begin{equation}\label{logArea}
\cl{1}{1},\cl{2}{0}:\ \log\m{B}=\log\lVert
\m{B}\rVert+\left(\text{artanh}\frac{\lVert
B^\prime\rVert}{b_0}\right) \frac{B^\prime}{\lVert
B^\prime\rVert},\quad (B^{\prime2}>0)\land(b_0\ne 0).
\end{equation}
To this formula we should add a free MV
$\m{F}=f_0+F^\prime=f_0+f_1\e{1}+f_2\e{2}+f_{12}\e{12}$ that
satisfies $\ee^{\m{F}}=1$, or
\begin{equation}
\ee^{\m{F}}=\ee^{f_0}\ee^{F^\prime}=\ee^{f_0}\bigl(\cosh\lVert
F^\prime\rVert+ \frac{F^\prime}{\lVert F^\prime\rVert}\sinh\lVert
F^\prime\rVert\bigr)=1.
\end{equation}
The solution of this MV equation (equivalently of four scalar
equations) is $f_0=0$, and $\lVert
F^\prime\rVert=\sqrt{f_1^2-f_2^2+f_{12}^2}=0$ for \cl{1}{1} and
$\lVert F^\prime\rVert=\sqrt{f_1^2+f_2^2-f_{12}^2}=0$ for
\cl{2}{0}. From this we conclude that $\m{F^\prime}=0$. Thus,  in
the case $B^2>0$ the principal logarithm is the only solution.

\subsection{\label{specificCases}\cl{1}{1} and \cl{2}{0} algebras: summary}

Taking into account the generic and special cases finally we can
write
\begin{align}\label{Log20Full}
& \cl{1}{1},\cl{2}{0}:\  \log(\m{B})=\notag\\
&\begin{cases}
  \log\lVert\m{B}\rVert+\bigl(\arctan (b_0,\lVert B^{\prime}\rVert)+2\pi k\bigr)\frac{B^{\prime}}{\lVert B^{\prime}\rVert},\qquad
   (B^{\prime 2}<0),\\
   \log \lVert\m{B}\rVert +
  \arctanh\bigl(\frac{\lVert B^{\prime}\rVert}{b_0}\bigr) \frac{B^{\prime}}{\lVert B^{\prime}\rVert},
  \qquad (B^{\prime 2}>0) \land (b_0 > 0)\land (b_0^2-B^{\prime 2}>0),\\
  \log(b_0)+\frac12 \log(0_{+})\bigl(1-\frac{B^{\prime}}{\lVert B^{\prime}\rVert}\bigr) +\frac12\log(2)\bigl(1+\frac{B^{\prime}}{\lVert B^{\prime}\rVert}\bigr) ,
  (B^{\prime 2}>0) \land (b_0 > 0)\land (\lVert\m{B}\lVert= 0),\\
  \log(b_0)+\frac{B^{\prime}}{b_0} ,
  \quad (B^{\prime 2}=0)\land (b_0>0),\\
  \log(-b_0)+(\pi+2\pi k)\hat{F}^{\prime} ,
  \quad (B^{\prime}= 0)\land (b_0\le 0),\\
  \varnothing,  \quad \textrm{no solution},\quad (B^{\prime 2}>
0)\land \bigl((b_0\le 0)\lor (b_0^2-B^{\prime 2} \le 0) \bigl).
\end{cases}
\end{align}
where $\lVert\m{B}\rVert=\sqrt{\abs(b_0^2-B^{\prime 2})}$ and
$\lVert B^{\prime}\rVert=\sqrt{\abs(B^{\prime 2})}$.
Explicit form of a free MV $\hat{F}^{\prime}$ is algebra dependent and is given in
\eqref{hatFCL11CL29}.

In conclusion, we shall remark that  in  \cl{0}{2} algebra the GA
logarithm is defined for all MVs. However, in the remaining 2D
algebras we have to satisfy the conditions for coefficients for a
logarithm to exist. Thus, in these algebras there are sectors in a
domain of argument where logarithm does not exist at all.

\vspace{3mm}
 \textbf{Example~9.} \cl{2}{0}, line~2 in
\eqref{Log20Full}. Case $(B^{\prime 2}>0) \land (b_0 > 0)\land
(b_0^2-B^{\prime 2}>0)$. \\ $\m{B}=2+B^\prime=2+5 \e{1}-\e{2}-5
\e{12}$; $B^{\prime 2}=1$, $\lVert\m{B}\rVert=\sqrt{3}$, $\lVert
B^\prime\rVert=1$, $\lVert\m{B}\rVert^2=b_0^2-B^{\prime 2}=3$.
\\Answer: $\log(\m{B})=\frac{\log (3)}{2}+\arctanh\left(\frac{1}{2}\right) (5
\e{1}-5 \e{12}-\e{2})$.

\vspace{1mm} \textbf{Example~10.}  \cl{2}{0}, line~3 in
\eqref{Log20Full}. Case $(B^{\prime 2}>0) \land (b_0 > 0)\land
(b_0^2-B^{\prime 2} = 0)$. \\$\m{B}=9-9 \e{1}+8 \e{2}+8 \e{12}$;
 $B^{\prime 2}=81$, $\lVert B^\prime\rVert=9$, $\lVert\m{B}\rVert^2=b_0^2-B^{\prime
2}=0$.
\\Answer: $\log(\m{B})= \frac{1}{18}\log\left(\frac{2}{0_{+}}\right) (-9 \e{1}+8 \e{2}+8 \e{12})+\frac{1}{2}
\log (2+0_{+})+\log (9)$, where $0_{+}$ is infinitesimally small
positive number. $\lim_{0_{+}\to 0}\ee^{\log\m{B}}=\m{B}$

\vspace{1mm} \textbf{Example 11}  \cl{2}{0}, line~4  in
\eqref{Log20Full}. Case $(\m{B}^{\prime 2}=0)\land (b_0>0)$\\
\quad $\m{B}=2+3\e{1}-4\e{2}-5\e{12}=2+B^\prime$,\quad$\lVert
\m{B}\rVert=2$,\quad$B^{\prime2}=0$, \quad$\lVert
B^\prime\rVert=0$. \\Answer: $\log\m{B}=\log 2+B^\prime/b_0=\log
2+\frac12(3\e{1}-4\e{2}-5\e{12})$.

\vspace{1mm} \textbf{Example 12.} \cl{2}{0}, line~5  in
\eqref{Log20Full}. Case $ (\m{B}^{\prime}= 0)\land (b_0\le 0)$.
The logarithm of MV $\m{B}=-2$ is $\log(\m{B})=\log (2)+(\pi +2
\pi k) \left(f_1 \e{1} -\e{12} \sqrt{1+f_1^2+f_2^2}+f_2
\e{2}\right)$. After exponentiation $\hat{F}^\prime$ simplifies
out and we get $\ee^{\log\m{B}}=-2$,

 \vspace{1mm}
 \textbf{Example~13.}  \cl{1}{1}, line~6 in~\eqref{Log20Full}
$\m{B}=2+5\e{1}-4\e{2}-7\e{12};$\quad$B^{\prime2}=58$, \quad$b_0^2-B^{\prime
2}=-54$. Logarithm does not exist, since under the condition $B^{\prime2}>0$ solution exists only when
$b_0^2-B^{\prime 2}>0$ (case~2 of \eqref{Log20Full}). Here we have $b_0^2-B^{\prime 2} <0$ and, therefore by line~6 of \eqref{Log20Full} the solution set is empty.

\vspace{4mm} The knowledge of logarithm and exponential provides a
possibility to calculate the  square root of a MV by formula
$\sqrt{\m{B}}=\pm\exp(\frac12\log\m{B})$. For example,  for
$\m{B}=2-\e{1}+2\e{12}$ in \cl{2}{0} we have $B^{\prime 2}=-3$ and
$\lVert\m{B}\rVert=\sqrt{7}$. Then,  using the
formula~\eqref{logCL11CL20} one finds $\log\m{B}=\frac12\log
7-\tfrac{1}{\sqrt{3}}(\e{1}-2\e{12})\arctan\tfrac{1}{\sqrt{3}}$
and after multiplication by $\tfrac{1}{2}$ and application of
exponential one obtains the root
$\sqrt{\m{B}}=\tfrac{2+\sqrt{7}-\e{1}+2\e{12}}{\sqrt{2(2+\sqrt{7})}}$.
In this way calculated root coincides with a root
formula~\eqref{sqrtroot} in the next subsection. It is easy to
verify that the geometric product $\sqrt{\m{B}}\sqrt{\m{B}}$
simplifies to the initial $\m{B}$. The formula
$\sqrt{\m{B}}=\pm\exp(\frac{1}{2}\log\m{B})$ gives only two
(plus/minus) roots. If $\m{B}$ is a unity, $\m{B}=1$, then the
square root exists for all algebras since
$\sqrt{1}=\pm\exp(\tfrac12\log (1))=\pm 1$. From
works~\cite{AcusDargysPreprint2020,Dargys2019} we know that in the
Clifford number algebra, in general, the square root of MV is a
multi-valued function. Below in Subsec.~\ref{sec:sqrt} such
(isolated) roots are presented for $\sqrt{+1}$ and $\sqrt{-1}$.
The multiple roots also have been found for a general MV in
\cl{1}{1} (see the next subsection) where up to four roots may
happen simultaneously. In higher dimensional GAs the number of
roots may be even larger~\cite{AcusDargysPreprint2020}.

\section{Addendum: Formulas for square roots of MV}\label{sec:sqrt}
Below the square roots $\sqrt{\m{B}}$ for 1D and 2D algebras that
may be useful in practice are presented. They were calculated from
the defining equation $\m{B}=\m{A}^2$ which is equivalent to a
system of two in 2D or four 2D real coupled equations.

For \cl{0}{1} there are two (plus and minus) roots of
$\m{B}=b_0+\e{1}b_1$,
\begin{equation}\label{sqrtCL01DE}
\cl{0}{1}:\sqrt\m{B}=\pm\frac{\sqrt{b_0^2+b_1^3}+(b_0+b_1\e{1})}
{\sqrt{2}\sqrt{b_0+\sqrt{b_0^2+b_1^2}}}=
\pm\frac{\magnitude{\m{B}}\m{B}}{\sqrt{2}\sqrt{\langle\m{B}\rangle_0+\magnitude{\m{B}}}},
\end{equation}
where $\magnitude{\m{B}}=\sqrt{\m{B}\gradeinverse{\m{B}}}$ is the
magnitude (norm) and $\langle\m{B}\rangle_0 =b_0$ is the scalar
part of MV. The roots~\eqref{sqrtCL01DE} exists for all MVs
$\m{B}\ne 0$.

For \cl{1}{0} algebra, in general, there are four roots of
$\m{B}=b_0+\e{1}b_1$,
\begin{equation}\label{CL10sqrtDE}
\cl{1}{0}:\sqrt{\m{B}}=\left\{\pm\frac{-\sqrt{b_0^2-b_1^2}+b_0+b_1\e{1}}{\sqrt{2}\sqrt{b_0-\sqrt{b_0^2-b_1^2}}},\
\pm\frac{\sqrt{b_0^2-b_1^2}+b_0+b_1\e{1}}{\sqrt{2}\sqrt{b_0+\sqrt{b_0^2-b_1^2}}}
\right\}.
\end{equation}
If  $b_0> b_1$ all roots are  different. If $b_0=b_1$ only two
distinct  roots remain.

In 2D algebras the MV is
$\m{B}=b_0+b_1\e{1}+b_2\e{2}+b_{12}\e{12}$.  The square root has
the same form for all algebras:
\begin{equation}\label{sqrtroot}
\cl{0}{2},
\cl{1}{1},\cl{2}{0}:\sqrt{\m{B}}=\pm\frac{b_0+\sqrt{\Det\m{B}}+b_1\e{1}+b_2\e{2}+b_{12}\e{12}}
{\sqrt{2}\sqrt{b_0+\sqrt{\Det\m{B}}}},
\end{equation}
For individual algebras the determinant is
\begin{equation}
\Det\m{B}=\m{B}\cliffordconjugate{\m{B}}=
\begin{cases}
&b_0^2+b_1^2+b_2^2+b_{12}^2\quad \text{for}\ \cl{0}{2},\\
&b_0^2-b_1^2+b_2^2-b_{12}^2\quad \text{for}\ \cl{1}{1},\\
&b_0^2-b_1^2-b_2^2+b_{12}^2\quad \text{for}\ \cl{2}{0}.\\
\end{cases}
\end{equation}
For Hamilton quaternion the square root exists for all MVs. For
\cl{1}{1} and \cl{2}{0} algebras the roots  exist when the
determinant is positive or zero. In  \cl{1}{1}  an additional
plus/minus roots may appear if
$\Det\m{B}>b_0>0$~\cite{Falcao2018},
\begin{equation}
\sqrt{\m{B}}=\pm\frac{-b_0+\sqrt{\Det\m{B}}-(b_1\e{1}+b_2\e{2}+b_{12}\e{12})}
{\sqrt{2}\sqrt{-b_0-\sqrt{\Det\m{B}}}}.
\end{equation}

\textit{Square roots of $+1$ in 2D}. The square roots of
$\m{B}=+1$ in 2D are
\begin{align}\label{sqrtpliua1}
\mkern-0mu& \sqrt{+1}=\begin{cases}
 \cl{0}{2}:\ \{\pm 1\},\\
\cl{1}{1}:\ \{\pm 1,\e{1},\tfrac{1}{\sqrt{2}}(\e{1}\pm\e{12})\} \textrm{ and } \pm(c_1 \e{1}+c_2\e{2}\pm\sqrt{1+c_1^2+c_2^2}\,\e{12}), \\
\cl{2}{0}:\ \{\pm
1,\e{1},\e{2},\tfrac{1}{\sqrt{2}}(\e{1}\pm\e{2})\}; \pm(c_1
\e{1}+c_2\e{2}\pm\sqrt{-1+c_1^2+c_2^2}\,\e{12}).\\
 \end{cases}
\end{align}
 In case of \cl{2}{0} and \cl{1}{1} the coefficients
$c_1$ and $c_2$ are arbitrary and may be considered as free
parameters. Their range is determined by expression under the
square root which must be positive. Thus, in this case we have two
types of roots: the isolated roots\index{square roots!isolated}
and the continuum of roots determined by  free parameters $c_i$.

\textit{Square roots of $-1$ in 2D}.\index{square
root!-1}\index{algebra!\cl{1}{1}}
\index{algebra!\cl{2}{0}}\index{algebra!\cl{1}{1}}\index{algebra!\cl{0}{2}}
The square roots of  $\m{B}=-1$ in 2D are
\begin{align}\label{sqrtminua1}
\mkern-0mu& \sqrt{-1}=
\begin{cases}
\cl{0}{2}:
\{\e{1},\e{2},\e{12},\tfrac{1}{\sqrt{2}}(\e{1}\pm\e{2}),\tfrac{1}{\sqrt{2}}(\e{1}\pm\e{12}),\tfrac{1}{\sqrt{2}}(\e{2}\pm\e{12})\}\\
\qquad\qquad\qquad\qquad\text{ and }\pm(c_1 \e{1}+c_2 \e{2}\pm\sqrt{1-c_1^2-c_2^2}\,\e{12}),\\
\cl{1}{1}:  \{\e{2}\}\text{ and }\pm(c_1 \e{1}+c_2 \e{2}\pm\sqrt{-1-c_1^2+c_2^2}\,\e{12}), \\
\cl{2}{0}: \{\e{12}\}\text{ and }\pm(c_1 \e{1}+c_2 \e{2}\pm\sqrt{1+c_1^2+c_2^2}\,\e{12}).\\
\end{cases}
\end{align}

\section{Conclusions and discussion} \label{sec:conclusion}

The paper presents exponential and logarithm functions of
multivector argument  for \cl{0}{1}, \cl{1}{0} (1-dimensional
commutative) and \cl{0}{2}, \cl{1}{1}, \cl{2}{0} (2-dimensional
non-commutative) Clifford geometric algebras (GAs). The well-known
approach to Hamilton quaternion identified  by  three imaginaries
$\{\ii,\jj,\kk\}$ was generalized and adapted, specifically,  the
imaginaries  have been  replaced by 2D unit multivectors the
squares of which are equal to~$\pm 1$ and which have been
constructed from Clifford basis vectors $\{\e{1},\e{2},\e{12}\}$.

The 2D basis-free  exponentials in this approach  assume a form of
either Euler's or de Moivre's  rules. The principal logarithm was
determined as an inverse of respective exponential. $2\pi$
multiplicity  in the logarithm was included by adding a free MV
$\m{F}$ that satisfies the condition $\ee^{\m{F}}=1$.  The
obtained formulas for exponential and logarithm  can be  applied
to quaternions too,  since \cl{0}{2}, \cl{1}{1} and \cl{2}{0}
algebras are isomorphic to, correspondingly, the Hamilton
quaternion, coquaternion, and conectorine~\cite{Opfer2017},

Since in GA the $n$-th root of MV is
$(\m{A})^{1/n}=\exp\big(\tfrac{1}{n}\log(\m{A})\big)$ the obtained
exponential and logarithm formulas may be applied to extract the
$n$-th root from general MV. However, this exp-log  formula allows
to calculate no more then two (plus/minus) square roots. Workable
formulas for square roots are presented. In particular, using the
defining equation we  have found explicit formulas for square
roots in 1D and 2D Clifford algebras and the sectors of their
existence in MV coefficient space.  In the space of MV
coefficients the sectors where the roots do not exist the
logarithm does not exist as well. Also, multiple square roots (up
to four) have been found in \cl{1}{0} and \cl{1}{1} algebras.

The presented results may be useful in applied GAs, especially in
dealing with GA differential equations the solutions of which are
expressed through GA
exponentials~\cite{AcusDargys2022a,DargysAcus2022a}. Finally, the
low dimensional Clifford algebras may be helpful in doing
calculations in higher dimensional algebras as well, because the
former are subalgebras of the latter. Also, it is expected that
the described in the paper approach may be adapted to higher grade
Clifford algebras.


%
\bibliographystyle{REPORT}
\enlargethispage{12pt}
\bibliography{exp-logdim1-2bNAMC}

\end{document}